\documentclass[twocolumn,showpacs,preprintnumbers,amsmath,amssymb,superscriptaddress]{revtex4-2}
\usepackage{graphicx}% Include figure files
\usepackage{dcolumn}% Align table columns on decimal point
\usepackage{bm}% bold math
\usepackage{float}
\usepackage{hyperref}
\usepackage{xcolor}
\usepackage{amsmath}
\hypersetup{
  colorlinks   = true, %Colours links instead of ugly boxes
  urlcolor     = red, %Colour for external hyperlinks
  linkcolor    = blue, %Colour of internal links
  citecolor   = blue %Colour of citations
}

\begin{document}
\graphicspath{{./figures/}}

\title{Methodology for Analyzing Proton Multiplicity Fluctuations with Azimuthal Partitions in Heavy-Ion Collisions}

\author{Dylan Neff}\email{dylan.neff@cea.fr}
\affiliation{Department of Physics and Astronomy, University of
  California, Los Angeles, California 90095, USA}
\affiliation{IRFU, CEA, Université Paris-Saclay, 91191 Gif-sur-Yvette, France}
\author{Zhongling Ji}
\affiliation{Department of Physics and Astronomy, University of
  California, Los Angeles, California 90095, USA}
\affiliation{Physics Department, Syracuse University, Syracuse, NY 13244, USA}
\author{Roli Esha}
\affiliation{Department of Physics and Astronomy, Stony Brook University, Stony Brook, NY 11790 USA} 
\author{Gang Wang}
\affiliation{Department of Physics and Astronomy, University of
  California, Los Angeles, California 90095, USA}
\author{Huan Zhong Huang} 
\affiliation{Department of Physics and
  Astronomy, University of California, Los Angeles, California 90095,
  USA} 
\affiliation{Key Laboratory of Nuclear Physics and Ion-beam
  Application (MOE), and Institute of Modern Physics, Fudan
  University, Shanghai-200433, People’s Republic of China}

\begin{abstract}
A primary objective in high-energy heavy-ion collisions is to investigate the phase transition between confined and deconfined color matter. Complementary to the cumulants of conserved charges integrated over the full azimuth,
%measured within a fixed rapidity window from collisions with a broad beam energy range, 
we introduce a novel experimental approach to explore particle fluctuations in azimuthal partitions, which are potentially sensitive to the first-order phase transition in heavy-ion collisions. We evaluate proton multiplicity ($N_w$) fluctuations in azimuthal partitions of width $w$ to quantitatively estimate the clustering tendency among these protons. The $\Delta \sigma^2$ observable is defined as the normalized difference between the variance of the $N_w$ distribution and the binomial baseline. We demonstrate the feasibility and characteristics of this observable through simulations using the AMPT and MUSIC+FIST models. We also use a Gaussian correlation model to illustrate that the dependence of $\Delta \sigma^2$ on $w$ can be parameterized to accurately extract the strength and the range of the input interaction among protons.
\end{abstract}

% \pacs{25.75.Ld}

\maketitle

% Table of contents only for organizational purposes while writing. Will be deleted once paper structure is fixed.
% \tableofcontents
% \vspace{1.5cm}
% Table of contents will be removed before publication.

\newpage

\section{Introduction}

Cumulants, particularly higher-order ones, of net-baryon distributions are expected to be sensitive to the order of the phase transition between the deconfined parton system and normal nuclear matter in high-energy heavy-ion collisions~\cite{net_proton_cumulants1, net_proton_cumulants2, net_proton_cumulants3}. In experiments, neutrons are typically undetectable, and the net-proton number is measured as a proxy for the net-baryon number. Considerable effort has been devoted to the measurements of proton and net-proton cumulants, as demonstrated, for instance, in Refs.~\cite{STAR:net_proton_kurtosis, STAR:C5_C6}.
Yet, the accurate measurement of cumulants in experiments and their comparison with theoretical predictions have both posed formidable difficulties.
Higher-order cumulants are not only more sensitive to the pertinent physics, but also more susceptible to detector anomalies, where a few problematic events can have an outsized impact~\cite{PhysRevC.109.034911}. Moreover, the statistical uncertainty in cumulant measurements rises with order, making higher-order analyses more challenging.
On the other hand, data interpretation must consider other effects, such as the finite volume and lifetime of the collision system~\cite{critical_slow_down}, initial volume fluctuations~\cite{PhysRevC.84.014904}, baryon number transport to midrapidities, and distinctions between net-proton and net-baryon cumulants~\cite{PhysRevC.85.021901}.   

In this paper, we propose an alternative approach to investigate proton fluctuations and to explore its possible sensitivity to the nature of the phase transition. During a hypothetical first-order phase transition, the surface tension between regions of different stable phases is anticipated to produce clumps of distinct baryon densities in the cooling medium~\cite{spinodal_density_enhancements}. These clumps in coordinate space could manifest in momentum space through the radial expansion of the system and, upon hadronization, may result in correlations among final-state particles.  Accordingly, we can investigate the potential increase in the clustering of final-state protons as a signal of the first-order phase transition, though the size of this signal may be significantly reduced when mapping from coordinate space to momentum space.

\begin{figure}[b]
\centering
\includegraphics[width=0.7\linewidth]{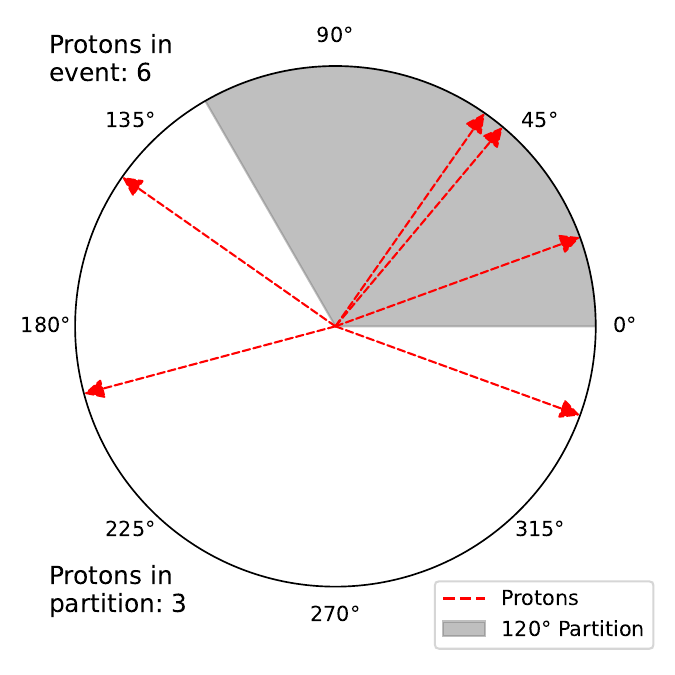}
\caption{Example event with six total protons, including three within a $120^\circ$-partition (shaded area). Red arrows represent the directions of proton momenta.}
\label{fig:az_partition_cartoon}
\end{figure}

\begin{figure}
\centering
\includegraphics[width=0.9\linewidth]{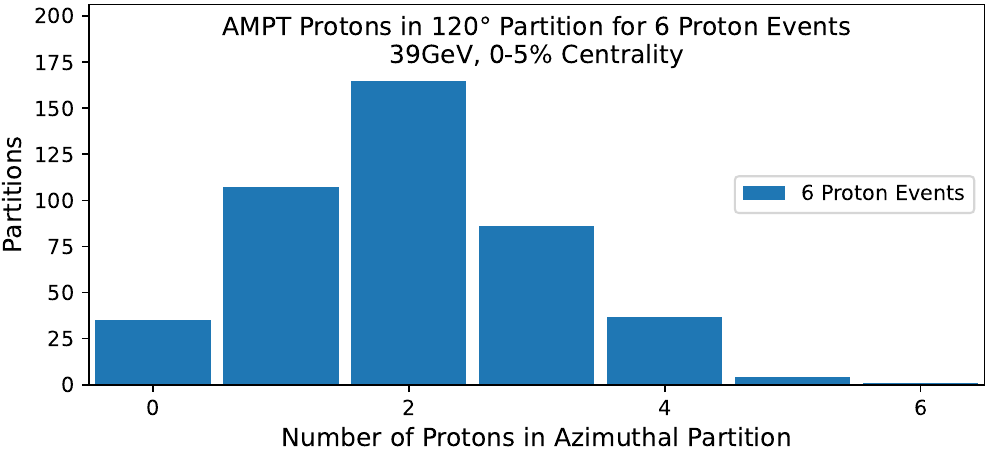}
\caption{Example $N_{120}$ distribution for 500 events with $N = 6$ in the 0--5\% centrality range of Au+Au collisions at $\sqrt{s_{NN}} = 39$ GeV from AMPT simulations.}
\label{fig:az_partition_example}
\end{figure}

To examine proton clustering, we analyze the event-by-event distribution of proton multiplicity ($N_w$) within an azimuthal partition of width $w$ (in degrees) randomly placed on each event.
Figure~\ref{fig:az_partition_cartoon} illustrates a mock event with $N_{120} = 3$, where the total proton multiplicity is $N = 6$. Figure~\ref{fig:az_partition_example} provides an example of the $N_{120}$ distribution for events with $N = 6$ in the 0--5\% centrality range of Au+Au collisions at $\sqrt{s_{NN}} = 39$ GeV from a multiphase transport (AMPT) model~\cite{ampt_original}.
If protons within an event are completely uncorrelated, the probability of a proton falling into a $w$-partition is $P_0 = \frac{w}{2\pi}$, and the $N_{w}$ values for events with $N$ protons follow a binomial distribution with $N$ trials and the probability of success $P_0$. 
To quantify proton clustering,
we define the $\Delta \sigma^2$ observable,
\begin{equation}
\Delta \sigma^2 \equiv \frac{\sigma^2(N_w) -  \sigma^2_{\text{binomial}}}{N (N-1)},
\label{eq:dsig2}
\end{equation}
where $\sigma^2(N_w)$ is the variance of the measured $N_{w}$ distribution, and the baseline $\sigma^2_{\text{binomial}}=NP_0(1-P_0)$ is the binomial variance expected from uncorrelated protons.
The normalization factor $N(N-1)$ will be explained in Sec.~\ref{sec:normalization}.

\begin{figure}
\centering
\includegraphics[width=\linewidth]{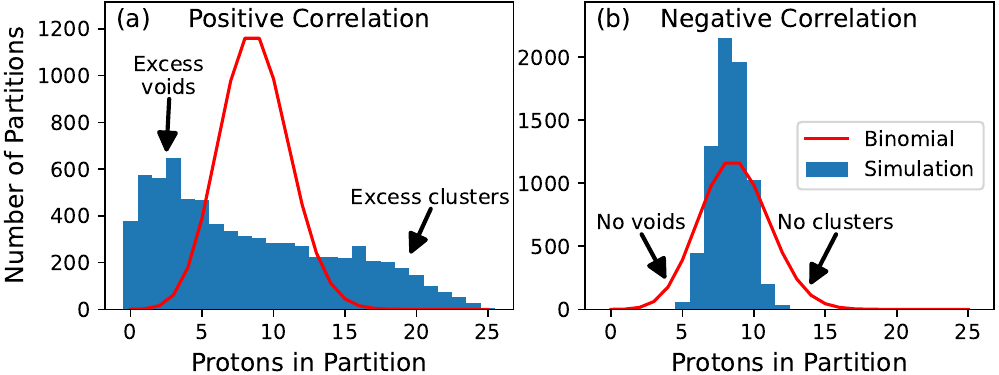}
\caption{
Simulations of the $N_w$ distributions with significant attraction (a) and repulsion (b) from a Gaussian correlation model discussed in Sec.~\ref{sec:gaus_cor_model}. Red lines represent the binomial distributions expected from uncorrelated protons.
}
\label{fig:clusters_voids}
\end{figure}

When an attractive interaction forms proton clusters in azimuth, corresponding voids will also emerge. Therefore, the randomly placed partitions will sample both excess clusters and excess voids, resulting in a $N_w$ distribution broader than the binomial reference and hence a positive $\Delta \sigma^2$.
Conversely,
a repulsive interaction tends to distribute protons more uniformly, leading to fewer clusters or voids, and producing a negative $\Delta \sigma^2$. 
Figure~\ref{fig:clusters_voids} depicts the $N_w$ distributions with significant attraction (a) and repulsion (b) from a Gaussian correlation simulation, displaying narrower and wider widths, respectively, than the binomial distribution (red lines).

The novel $\Delta \sigma^2$ observable possesses several strengths that we aim to leverage. The normalization scheme renders it insensitive to homogeneous detector inefficiencies, while an event-mixing technique mitigates contributions from azimuthal-dependent detector inefficiencies and other detector effects. Furthermore, multiple sampling over the same event (event resampling) optimally utilizes the statistics of the data.
More importantly, beyond technical aspects, azimuthal partitions unveil a finer structure of the collision event than an integrated conserved charge.

In Sec.~\ref{sec2}, we outline the comprehensive methodology for measuring proton $\Delta \sigma^2$. %Subsequently, we conduct an analytical study of $\Delta\sigma^2$ in Sec.~\ref{sec:analytical_regime}. 
We proceed to study the dependence of this observable on $N$ and $w$ in simulations using the AMPT and MUSIC+FIST models~\cite{music1,music2,music3,music4,music5,fist_original} in Sec.~\ref{sec:model_analysis}. In Sec.~\ref{sec:gaus_cor_model}, we develop a Gaussian correlation model to examine the observable's sensitivity to the strength and the range of proton correlations.  
A summary and outlook are provided in Sec.~\ref{sec6}. Additional details and tests on momentum conservation are included in Appendix~\ref{sec:mom_cons_model}.

\section{Methodology}
\label{sec2}

In the subsequent analyses, we select the kinematic region for protons similarly to the RHIC Beam Energy Scan phase I (BES-I) data, with $|y|<0.5$ and $0.4 < p_T < 2$ GeV/$c$. 
Events are divided into separate classes based on $N$ and collision centrality, and a $N_w$ distribution is constructed for each class. In this section, we will address the technical details related to normalization, event mixing, correction for elliptic flow, and event resampling.

\subsection{\texorpdfstring{ $\Delta \sigma^2$ }{}Observable}
\label{sec:normalization}

We begin with $N$ protons, each independently distributed in azimuth, with the probability density function (PDF) $p(\varphi)$. The integrated probability for each proton falling in the azimuthal window $[\psi, \psi+w]$ is $P(\psi)=\int_{\psi}^{\psi+w} p(\varphi) d\varphi$. Here, $\psi$ represents the orientation of the PDF, which we take to be evenly distributed. The variance of the $N_w$ distribution is
\begin{equation}
\begin{split}
\sigma^2(N_w) =& E_{\psi}\left[ E_{N_w|\psi}[N_w^2] \right] - (E_{\psi}\left[ E_{N_w|\psi}[N_w] \right])^2 \\
=& E_{\psi}\left[ NP(\psi)\left(1 - P(\psi)\right) + \left(NP(\psi)\right)^2 \right] - (NP_0)^2 \\
=& NP_0(1-P_0) + N(N-1)\left(E_{\psi}[P^2(\psi)]-P_0^2\right) \\
=& \sigma_{\text{binomial}}^2 + N(N - 1) \left[ \int_0^{2\pi} \frac{P^2(\psi)}{2 \pi} \,d\psi - P_0^2 \right],
\end{split}
\label{eq:var_final}
\end{equation}
where $\sigma^2_{\text{binomial}} \equiv NP_0(1-P_0)$ is the binomial variance for a uniform $p(\varphi)$ with $P_0 \equiv \frac{w}{2\pi}$, and $E_{X|Y}\left[f(X)\right] \equiv \sum_X P(X|Y) f(X)$ denotes the conditional expectation.
Rearranging Eq. (\ref{eq:var_final}), we can express the observable $\Delta \sigma^2$ from Eq.~(\ref{eq:dsig2}) as the normalized deviation from the binomial baseline: 
\begin{eqnarray}
\Delta \sigma^2 &\equiv& \frac{\sigma^2(N_w) - \sigma_{\text{binomial}}^2}{N(N - 1)} \label{eq:delta_var_def} \nonumber\\ &=& \int_0^{2\pi} \frac{P^2(\psi)}{2 \pi} \,d\psi - P_0^2. \label{eq:delta_var_def2}   
\end{eqnarray}
While the derivation is based on independent protons that only produce a non-negative $\Delta \sigma^2$~\cite{dylan_thesis}, we show in Sec.~\ref{sec:gaus_cor_model} that this observable is sensitive to proton interactions that could yield both positive and negative correlations.

\subsection{Event Mixing}
\label{sec:event_mixing}

Mixed events are commonly employed in heavy-ion collision analyses, as they account for detector effects present in all events while eliminating any intra-event correlation between particles. In particular, mixed events reflect azimuthal-dependent inefficiencies, whose contribution to $\Delta \sigma^2$ must be subtracted from same events,
\begin{equation}
\Delta \sigma^2_{\text{mix-corrected}} = \Delta \sigma^2_{\text{same}} - \Delta \sigma^2_{\text{mixed}}.
\label{eq:mixed_correction}
\end{equation}
While not vital to the model studies in this paper, which do not include detector effects, the event-mixing procedure is crucial in real-data analyses. In such analyses, a mixed event with $N$ protons can be generated by sampling each proton from a distinct event. The sampling is performed $N$ times from events with collision location and collision centrality
similar to the current event under study.

\subsection{Elliptic Flow Correction}
\label{sec:flow_correction}

In heavy-ion collisions, the geometric anisotropy of the initial overlap region can be transformed through hydrodynamic expansion~\cite{doi:10.1146/annurev-nucl-102212-170540} into the anisotropy of particle emission in momentum space. These collective motions mimic the clustering behavior in the $\Delta \sigma^2$ observable and, consequently, contribute as background to the physics related to the phase transition. In a specific kinematic region, the collective motions are usually quantified by decomposing
the $\varphi$ distribution of produced particles in each collision with a Fourier series~\cite{PhysRevC.58.1671}:
\begin{equation}
\frac{1}{N}\frac{dN}{d\varphi} =\frac{1}{2\pi} \big(1+ \sum_{n=1}^{\infty} 2v_n\cos n\Delta \varphi\big) ,
\label{Eq: Fourier expansion}
\end{equation}
where $\Delta \varphi$ is the azimuthal angle of a particle relative to the reaction plane (spanned by impact parameter and beam momenta).
The coefficients $v_n\equiv\langle \cos n\Delta\varphi\rangle$ are experimentally obtainable by averaging over particles of interest and over events.
Conventionally, we refer to $v_1$ as directed flow, $v_2$ as elliptic flow, and $v_3$ as triangular flow.

Applying the $p(\varphi)$ distribution in Eq.~(\ref{Eq: Fourier expansion}), we obtain the $v_n$ contribution to $\Delta \sigma^2$ in  Eq.~(\ref{eq:delta_var_def2}),
\begin{equation}
\Delta \sigma^2 _{v_n} = \frac{2 v_n^2}{\pi^2 n^2} \sin^2 \frac{nw}{2} .
\label{eq:flow}
\end{equation}
Directed flow is typically small in magnitude at midrapidities and exhibits an odd function of rapidity, allowing us to neglect its contribution to a symmetric rapidity region. The factor of $n^2$ in the denominator of Eq.~(\ref{eq:flow}) rapidly diminishes the contribution from higher-order harmonics, irrespective of their $v_n$ magnitudes. 
Therefore, we will only focus on the correction for
elliptic flow,
%We measure the $v_2$ values of our protons of interest (shown in Appendix~\ref{sec:v2_measurement}) 
and subtract the corresponding contribution according to Eq.~(\ref{eq:flow}). Note that the correction depends on $w$ and that the contribution vanishes for partitions with $w=180^\circ$.

\subsection{Event Resampling}

Instead of randomly placing a single azimuthal partition on each event, we can place multiple partitions on a single event and append these multiplicities to the $N_w$ distribution. This procedure, referred to as ``event resampling", optimally utilizes the statistics in each event at the cost of generating distributions composed of entries that are not fully independent.
The resampling is grounded on the assumption that, after correcting for flow and detector effects, the relevant physics has an equal probability in all azimuthal directions, regardless of the scale of $w$. In other words, there is no preferred azimuthal direction for a potential phase transition signal.
Event resampling converges to the true moments of a distribution faster than taking a single sample (partition), and according to simulations in Ref.~\cite{dylan_thesis}, a Poisson block bootstrap procedure offers accurate estimates of the statistical uncertainties on the moments. We will, therefore, default to sampling 72 partitions per event in the following analyses.

\section{Application to AMPT and MUSIC+FIST Models}
\label{sec:model_analysis}

In this section, we highlight certain features of the $\Delta \sigma^2$ observable using simulated data from the AMPT and MUSIC+FIST models.
As $\Delta \sigma^2$ involves two independent variables, $N$ and $w$, we delve into the  dependence of $\Delta \sigma^2$ on each of them.

\subsection{Model Description}
\label{sec3a}
The AMPT model~\cite{ampt_original} is an event generator that simulates the microscopic interactions in relativistic heavy-ion collisions. The initial conditions are managed by the HIJING model~\cite{hijing1,hijing2}, with the creation of hadrons, strings, and minijet partons from initial scattering. We employ the string melting version of this model~\cite{ampt_string_melting}, in which the initial-state hadrons and minijets are converted into partons, whose interactions are described by Zhang's parton cascade (ZPC)~\cite{zhang_parton_cascade}. At freeze-out, hadronization is governed by a quark coalescence model, and a relativistic transport (ART) model~\cite{PhysRevC.52.2037} is then exploited to process the final-state hadronic scattering. 
The string melting version of AMPT reasonably well reproduces particle spectra and elliptic flow
in heavy-ion collisions at RHIC and the LHC~\cite{PhysRevC.90.014904}.
A recent update of AMPT with an improved quark coalescence algorithm~\cite{ampt_sm_improved} is used in this analysis.

MUSIC+FIST is a combination of the MUSIC hydrodynamic model~\cite{music1,music2,music3,music4,music5} and the FIST sampler~\cite{fist_original}. The collision system takes the initial conditions from the Monte Carlo Glauber model~\cite{glauber} and undergoes the hydrodynamic evolution in MUSIC. The medium expands until reaching a ``switching" energy density, $\varepsilon_{\text{SW}}=0.26$  GeV/fm$^3$. The FIST sampler then generates final-state particles based on the produced freeze-out hypersurface. This model pre-determines the number of each particle species to sample in each event, ensuring global conservation of conserved quantum numbers, but at the cost of not conserving momentum. An excluded-volume effect has been incorporated into the original FIST model~\cite{fist_ev}, ensuring that upon sampling the hypersurface, no two baryons are sampled within a volume of radius $r=1$ fm. This radius is an adjustable parameter.

%An example of an azimuthal partition multiplicity distribution in AMPT data is shown in Figure~\ref{fig:example_dist} for 120$^\circ$ partitions of events containing 20 total protons in 0-5\% most central 39 GeV data. Superimposed in red circles is the expected binomial distribution with $N=20$ and $p=\frac{1}{3}$. The AMPT and binomial distributions look virtually identical, implying that the protons in AMPT exhibit little correlation. On closer inspection, one may be able to discern slight differences between the two distributions. Specifically, the AMPT distribution is slightly higher than the binomial expectation near the center of the distribution while slightly lower in the tails. From the discussion around Figure~\ref{fig:clusters_voids}, we can interpret this to mean that AMPT has fewer extreme partition multiplicities, implying less clustering than would be expected to occur randomly.

\begin{figure}
\includegraphics[width=\linewidth]{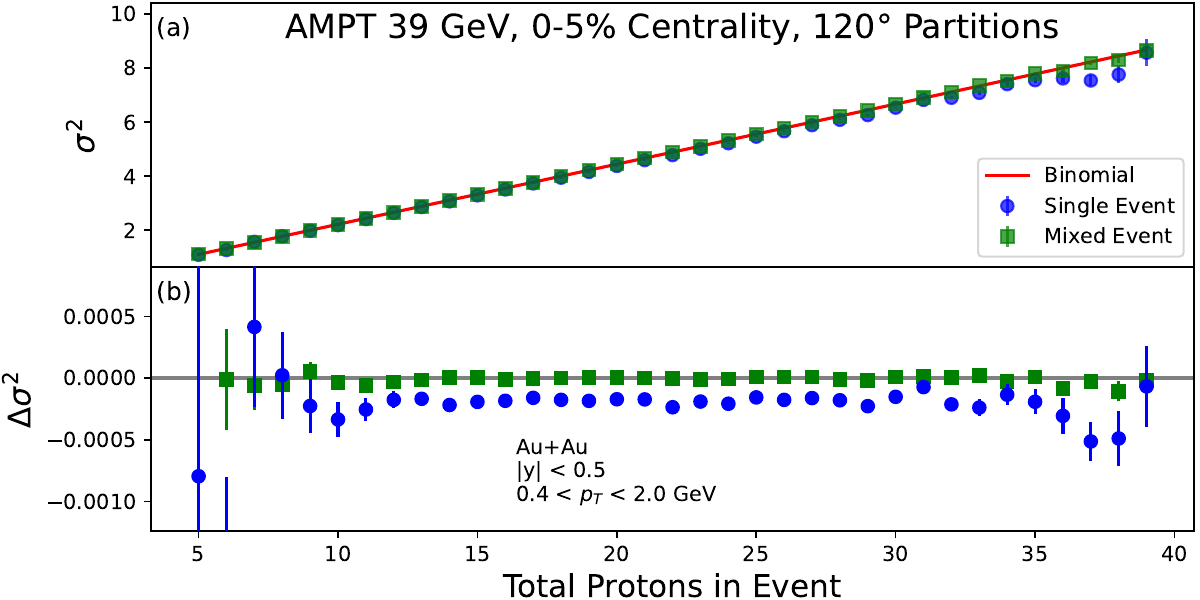}
\caption{(a) AMPT simulations of $\sigma^2(N_{120})$ vs $N$ in the 0--5\% most central Au+Au collisions at $\sqrt{s_{NN}} = $ 39 GeV using 72 partition samples per event, for both same events (blue circles) and mixed events (green squares). The red line corresponds to the binomial variance, $NP_0(1-P_0)$. (b) $\Delta \sigma^2$ vs $N$, calculated from $\sigma^2(N_{120})$ in panel (a) by subtracting the binomial variance and then dividing by $N(N-1)$.}
\label{fig:sig2_dsig2_example}
\end{figure}

%\begin{figure}
%\includegraphics[width=\linewidth]{figures/ampt_fist_analysis/variance_vs_tprotons.pdf}
%\caption{The variance of AMPT azimuthal partition multiplicity distributions at 39 GeV for the 0-5\% most central events using 120$^\circ$ partitions and 72 partition samples per event is shown for single events in blue and mixed events in green. Each point corresponds to the variance of a distribution constructed from events with the same number of protons within acceptance, $N$, represented on the x-axis. The red line corresponds to the binomial variance $Np(1-p)$.}
%\label{fig:sig2_example}
%\end{figure}

%\begin{figure}[b]
%includegraphics[width=\linewidth]{figures/ampt_fist_analysis/dsig2_vs_tprotons_ampt_39gev.pdf}
%\caption{The $\Delta \sigma^2$ observable is calculated from the variances in Figure~\ref{fig:sig2_example} by subtracting the binomial variance and then dividing by $N(N-1)$.}
%\label{fig:dsig2_example}
%\end{figure}

\begin{figure*}[bthp]
\includegraphics[width=\textwidth]{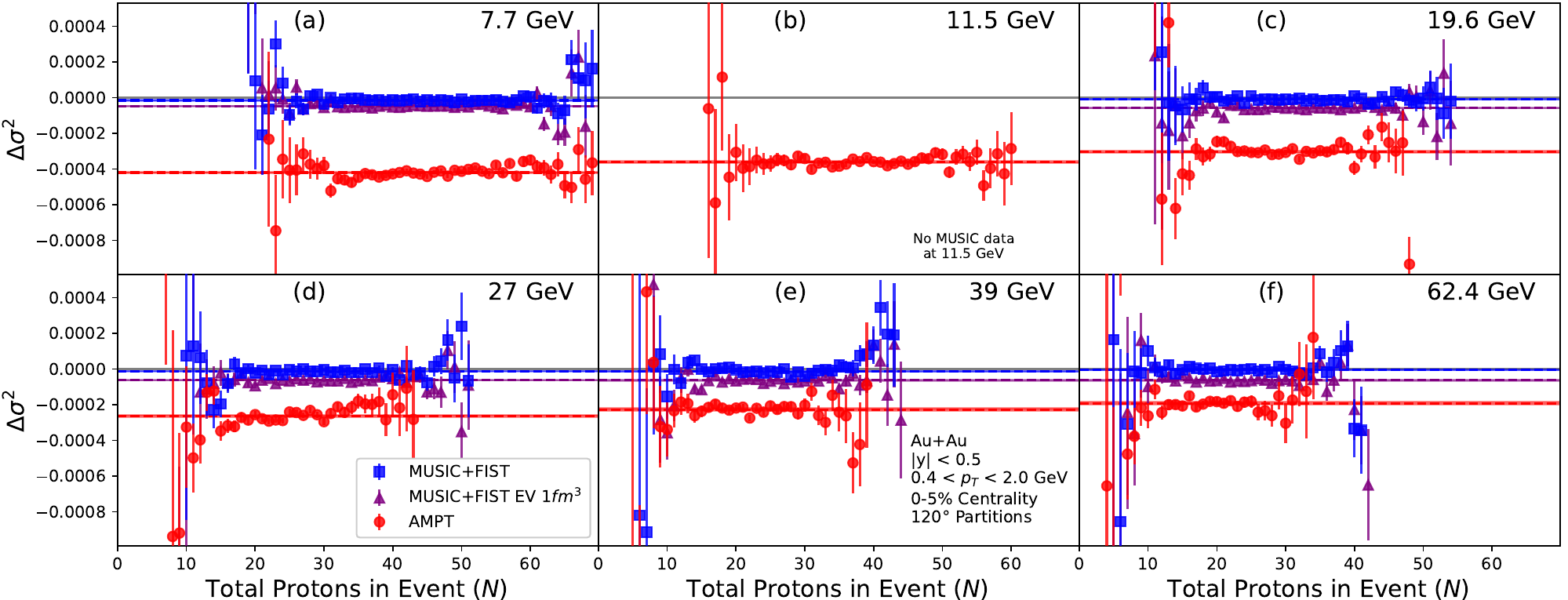}
\caption{$\Delta \sigma^2(N_{120})$ vs $N$ in 0--5\% Au+Au collisions at different beam energies from AMPT (red circles), default MUSIC+FIST (blue squares), and MUSIC+FIST with an excluded-volume effect (purple triangles). The horizontal dashed lines represent the weighted averages of each data set, with the bands about the lines representing statistical uncertainties.}
\label{fig:dsig2_vs_tprotons_energies}
\end{figure*}

\begin{figure}[b]
\includegraphics[width=\linewidth]{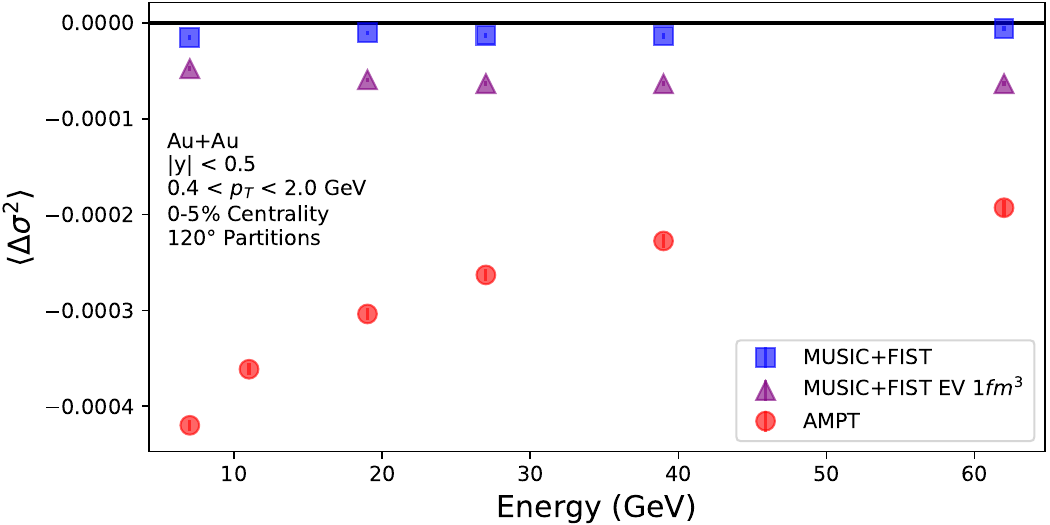}
\caption{$\langle \Delta \sigma^2 \rangle$, obtained by averaging the $\Delta \sigma^2$ values from Fig.~\ref{fig:dsig2_vs_tprotons_energies} over $N$, as a function of beam energy, from AMPT (red circles), default MUSIC+FIST (blue squares), and MUSIC+FIST with an excluded-volume effect (purple triangles).}
\label{fig:dsig2_vs_energy}
\end{figure}

\begin{figure}[b]
\includegraphics[width=\linewidth]{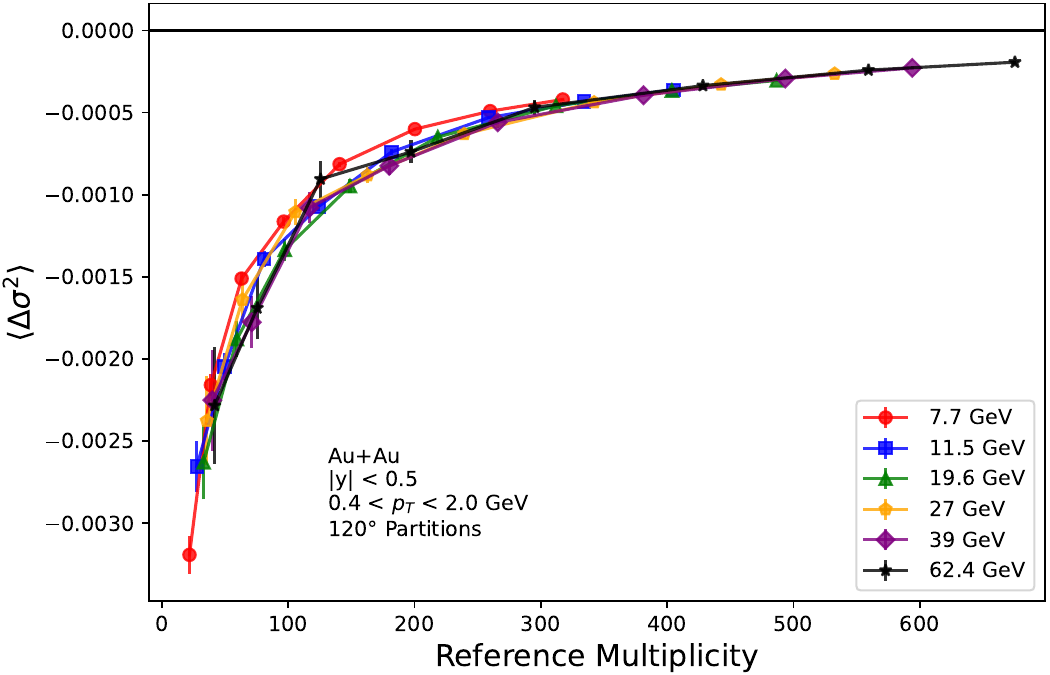}
\caption{AMPT simulations of $\langle \Delta \sigma^2 \rangle$ as a function of the average reference multiplicity for each centrality class in Au+Au collisions at various beam energies. Results for each of the six energies are plotted in a different color and connected by straight lines.}
\label{fig:dsig2avg_vs_refmult}
\end{figure}

\subsection{\texorpdfstring{$N$}{N} Dependence}

Figure~\ref{fig:sig2_dsig2_example}(a) shows the AMPT simulations of $\sigma^2(N_{120})$ vs $N$ in the 0--5\% most central Au+Au collisions at $\sqrt{s_{NN}} = $ 39 GeV using 72 partition samples per event, for both same events (blue  circles) and mixed events (green squares). 
The expected binomial variance (red line) $NP_0(1-P_0)$ is closely followed by
the mixed-event variance, slightly above the same-event one.
Figure~\ref{fig:sig2_dsig2_example}(b) presents the corresponding 
$\Delta \sigma^2$ values as a function of $N$.
$\Delta \sigma^2_{\text{mixed}}$ is consistent with zero, as AMPT lacks azimuthal inhomogeneities across events, unlike real data with detector effects. On the other hand, $\Delta \sigma^2_{\rm same}$ is significantly negative, implying an effectively repulsive interaction among protons in the AMPT events. 
In the subsequent analyses, we skip the correction with mixed events, and only correct $\Delta \sigma^2_{\rm same}$ for elliptic flow, as described in Section~\ref{sec:flow_correction}, to obtain the final observable $\Delta \sigma^2$.

%To better represent the systematic deviations of AMPT data from the binomial expectation, we calculate $\Delta \sigma^2$ from Equation~\ref{eq:dsig2} and plot this in Figure~\ref{fig:dsig2_example} vs $N$. A positive $\Delta \sigma^2$ indicates an azimuthal partition multiplicity distribution wider than binomial which would suggest the presence of clustering. A negative $\Delta \sigma^2$ indicates a distribution narrower than binomial and is indicative of a repulsive interaction among proton tracks. $\Delta \sigma^2=0$ indicates distributions of the same width as binomial, consistent with no correlation within the sensitivity of the observable. As there are fewer events with very large or very small $N$, the points on the far left and far right of this plot have lower statistics, which is reflected in their relatively large error bars.

\begin{figure*}
\includegraphics[width=\textwidth]{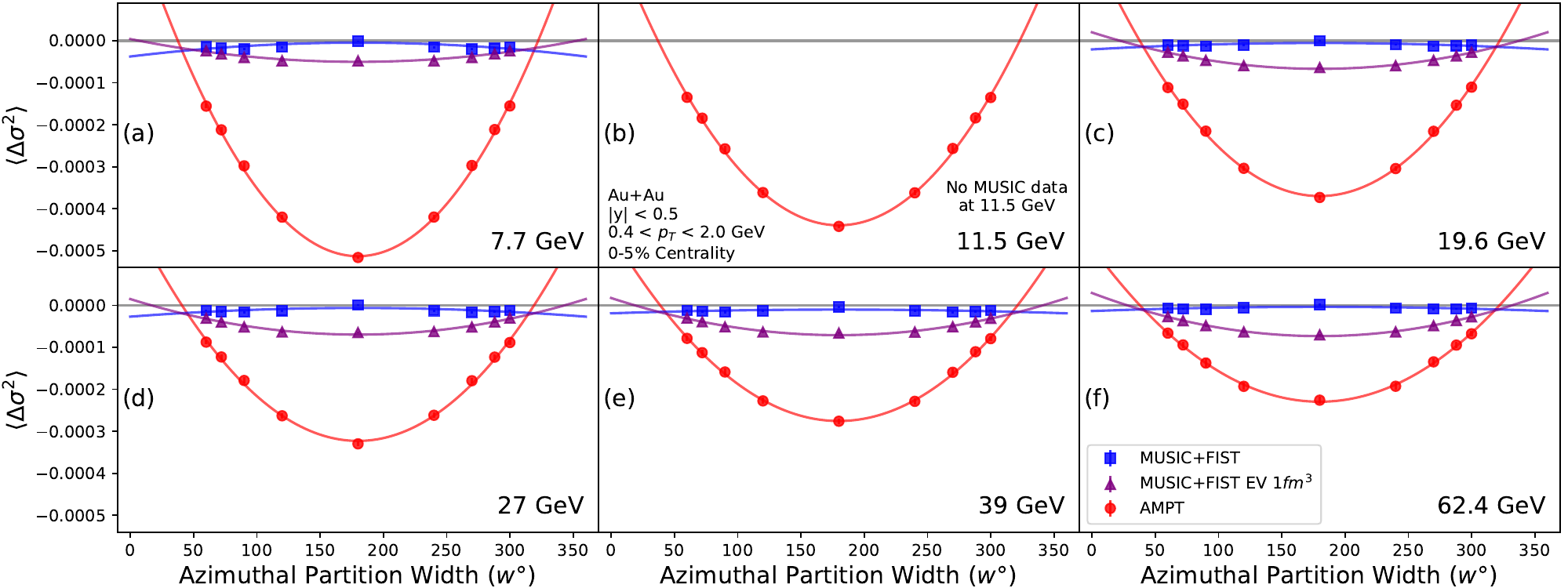}
\caption{$\langle \Delta \sigma^2 \rangle$ as a function of
$w$ in 0--5\% Au+Au collisions at various beam energies from AMPT (red), default MUSIC+FIST (blue), and MUSIC+FIST with an excluded-volume effect (purple). The points are fit to a quadratic function described in Eq.~(\ref{eq:quad_fit_ampt}).}
\label{fig:dsig2avg_vs_width}
\end{figure*}

%The lack of correlation in mixed events is found in general for both the AMPT and MUSIC+FIST model. \blue{I do the mixed correction in model which slightly increases the error bars, though maybe the following discussion can be shortened?} While event mixing will be important for correcting detector inhomogeneities in experimental data as described in Section~\ref{sec:event_mixing}, it will not be necessary in simulated data from these models. However, in order to mimic the analysis performed on experimental data, we will perform the mixed event correction on the simulated model data in this paper, according to Eq.~(\ref{eq:mixed_correction}).

%This correction is close to zero for the model data, but will slightly increase the statistical error bars due to the slight non-linearity of the $\Delta \sigma^2$ observable~\cite{dylan_thesis}. We will therefore perform the mixed correction by default and $\Delta \sigma^2$ will be assumed to be mix-corrected (the ``mix-corrected'' subscript will be suppressed). A correction for elliptic flow, as described in Section~\ref{sec:flow_correction}, is also made in the analysis.

Figure~\ref{fig:dsig2_vs_tprotons_energies} shows the flow-corrected $\Delta \sigma^2(N_{120})$ vs $N$ in 0--5\% Au+Au collisions at various beam energies from AMPT (red circles), default MUSIC+FIST (blue squares), and MUSIC+FIST with an excluded-volume effect (purple triangles). At all energies, the AMPT results are significantly negative, indicating repulsive interactions. The $\Delta \sigma^2$ values from both versions of MUSIC+FIST are much closer to zero, implying a weaker correlation between protons than in AMPT. In each data set, $\Delta \sigma^2$ remains roughly constant with respect to $N$, supporting the normalization scheme in Eq.~(\ref{eq:delta_var_def}), which is defined such that $\Delta \sigma^2$ should be independent of $N$. This constancy justifies averaging $\Delta \sigma^2$ over $N$, as represented by the horizontal lines.

Figure~\ref{fig:dsig2_vs_energy} shows
$\langle \Delta \sigma^2 \rangle$, obtained by averaging the $\Delta \sigma^2$ values from Fig.~\ref{fig:dsig2_vs_tprotons_energies} over $N$, as a function of beam energy. In the default MUSIC+FIST model, $\langle \Delta \sigma^2 \rangle$ is minimally but significantly negative, with no evident energy dependence. The excluded-volume version of MUSIC+FIST exhibits markedly more negative outcomes compared with the default version. Since the excluded-volume effect is expected to produce an effectively repulsive interaction, the more negative $\langle\Delta \sigma^2\rangle$ values corroborate that our observable is indeed sensitive to repulsion. 
The AMPT results seem to reveal much stronger repulsive interactions than MUSIC+FIST, and the repulsion strength intensifies at lower collision energies.

To explore the disparity between MUSIC+FIST and AMPT, as well as the beam-energy dependence of the latter, we depict in Fig.~\ref{fig:dsig2avg_vs_refmult} $\langle \Delta \sigma^2 \rangle$ as a function of the average reference multiplicity in each centrality class for the investigated beam energies. Reference multiplicity is defined by the number of charged hadrons within the pseudorapidity range of $|\eta|<0.5$ in an event.  The AMPT points at all energies appear to follow a universal curve, with the potential exception of the 7.7 GeV data being slightly less negative than other energies. This decreasing trend toward lower multiplicities explains the dependence on center-of-mass energy in Fig.~\ref{fig:dsig2_vs_energy}, because for the fixed 0--5\% centrality, collisions at lower energies produce fewer final-state particles.
We attribute this pronounced multiplicity dependence primarily to global momentum conservation. In a peripheral collision, where only a few particles are produced, each particle bears greater responsibility for conserving momentum. This tendency results in a more uniform distribution of particles in azimuth and, consequently, a negative $\Delta\sigma^2$. Each additional particle introduces an extra degree of freedom, allowing for a more flexible means to satisfy global momentum conservation, and the negative correlation is diluted. This effect approaches zero as more particles are added to the event. As mentioned in Sec.~\ref{sec3a}, the FIST sampler does not conserve momentum. Therefore, the MUSIC+FIST results do not manifest the same trend as AMPT in Fig.~\ref{fig:dsig2_vs_energy}. %\blue{MUSIC+FIST also only produced at 0-5\% most central, otherwise would certainly plot that data vs refmult as well. Don't know if this is worth mentioning here?}

This hypothesis of momentum conservation is substantiated by simulations presented in Appendix~\ref{sec:mom_cons_model}. In these simulations, particles are generated with random momenta which are slightly rotated until the total momentum of the event approaches zero. The resulting $\langle \Delta \sigma ^2 \rangle$ as a function of $M$, the total number of particles in the event, mirrors the trend observed in AMPT. 

\subsection{\texorpdfstring{$w$}{w} Dependence}

The preceding discussion has been confined to a partition width of $w=120^\circ$. Intuitively, varying the partition width may offer sensitivity to the dynamical scale associated with the measured correlation.
We will
investigate this sensitivity
in Sec.~\ref{sec:gaus_cor_model}
with a Gaussian correlation model. In this subsection, we first illustrate the dependence of $\langle \Delta \sigma^2 \rangle$ on $w$ and then discuss its parametrization.

Figure~\ref{fig:dsig2avg_vs_width} shows $\langle \Delta \sigma^2 \rangle$ as a function of
$w$ in 0--5\% Au+Au collisions at various beam energies from AMPT (red circles), default MUSIC+FIST (blue squares), and MUSIC+FIST with an excluded-volume effect (purple triangles). 
The apparent symmetry about $180^\circ$ arises from the mathematical nature of Eq.~(\ref{eq:delta_var_def2}). Substituting $w$ with $(2\pi-w)$ is equivalent to replacing $P_0$ with $(1-P_0)$ and, in the integral, replacing $P(\psi)$ with $[1-P(\psi)]$. Under this operation, Eq.~(\ref{eq:delta_var_def2}) remains unchanged. The proof entails the following identity,
\begin{eqnarray}
\int_0^{2\pi}  P(\psi) d\psi &=& \int_0^{2\pi}   d\psi \int_\psi^{\psi+w} p(\varphi) d\varphi \nonumber \\
&=& \int_0^w d\varphi \int_0^{2\pi} p(\varphi+\psi)d\psi \nonumber \\ 
&=& \int_0^w d\varphi = w.
\end{eqnarray}

\begin{figure}[b]
\includegraphics[width=\linewidth]{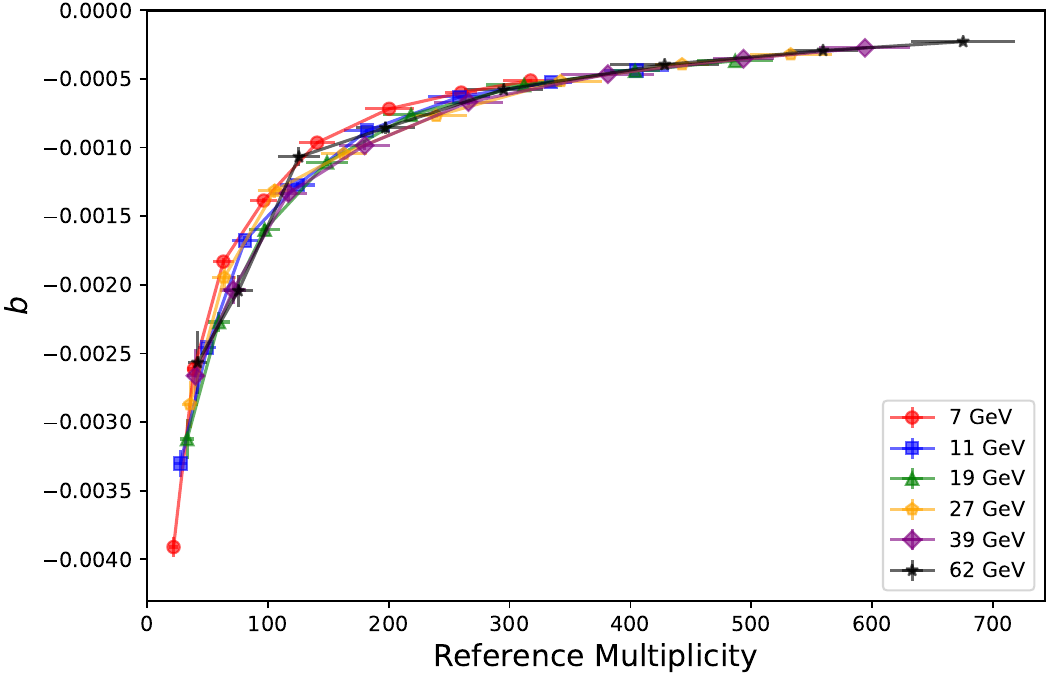}
\caption{The general scale $b$ extracted from the $w$ dependence of $\langle \Delta \sigma^2 \rangle$ in Fig.~\ref{fig:dsig2avg_vs_width} using the quadratic fit in Eq.~(\ref{eq:quad_fit_ampt})  as a function of reference multiplicity from AMPT.}
\label{fig:b_vs_refmult}
\end{figure}

\begin{figure}[b]
\includegraphics[width=\linewidth]{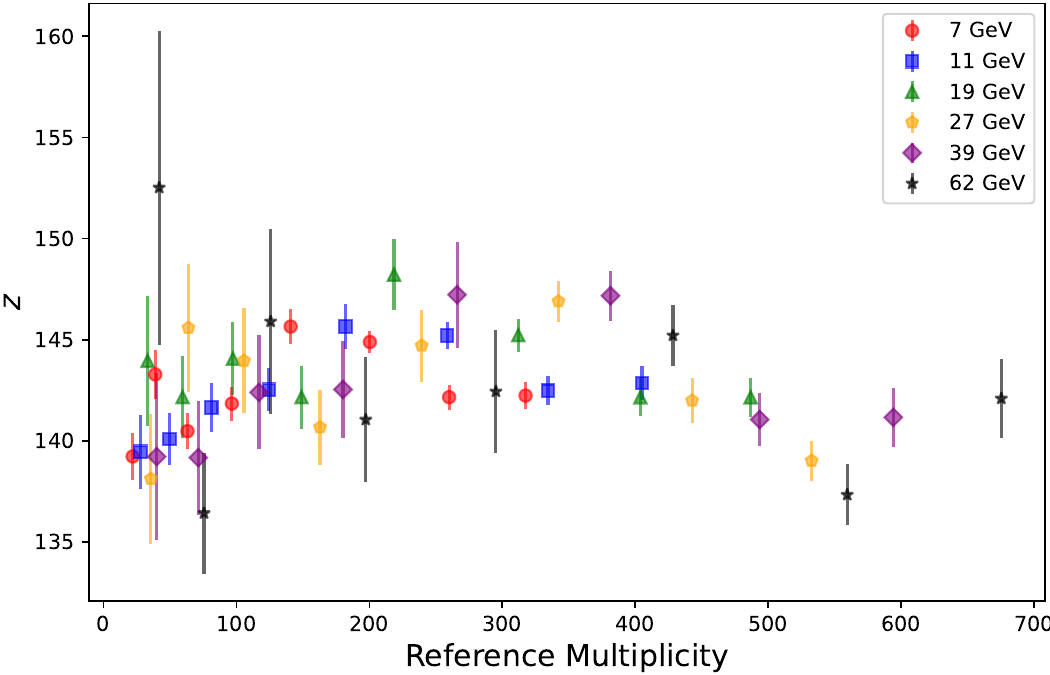}
\caption{The inverse curvature $z$ extracted from the $w$ dependence of $\langle \Delta \sigma^2 \rangle$ in Fig.~\ref{fig:dsig2avg_vs_width} using the quadratic fit in Eq.~(\ref{eq:quad_fit_ampt})  as a function of reference multiplicity from AMPT.}
\label{fig:z_vs_refmult}
\end{figure}

We parameterize the $w$ dependence of $\langle \Delta\sigma^2 \rangle$ with a quadratic function,
\begin{equation}
\langle \Delta\sigma^2 \rangle = b \left[1 - \left(\frac{w - 180}{z}\right)^2 \right],
\label{eq:quad_fit_ampt}
\end{equation}
where $b$ and $z$ quantify the general scale and the inverse curvature, respectively, supposedly reflecting the strength and the range
of the correlation between protons.
We extract these two parameters from the AMPT simulations in Fig.~\ref{fig:dsig2avg_vs_width}, and present $b$ and $z$ as a function of reference multiplicity in Figs.~\ref{fig:b_vs_refmult} and \ref{fig:z_vs_refmult}, respectively.
The universal trend in Fig.~\ref{fig:b_vs_refmult} looks akin to that in Fig.~\ref{fig:dsig2avg_vs_refmult}, with a larger scale, since
$b$ corresponds to the maximum $\langle \Delta \sigma^2 \rangle$ magnitude occurring at $w=180^\circ$.
Representing the general scale with $b$ removes the arbitrariness of the choice in $w$.
In Fig.~\ref{fig:z_vs_refmult}, there is no apparent dependence of $z$ on either beam energy or reference multiplicity.
We anticipate applying this methodology to real data and comparing the outcomes with model simulations to deepen our understanding of the pertinent physics.

\section{Analysis with Gaussian Correlation Model}
\label{sec:gaus_cor_model}

In this section, we examine the sensitivity of the $\Delta\sigma^2$ observable to proton interactions.
We devise a simple toy model incorporating both positive and negative Gaussian-type correlations among protons. We will demonstrate that our proposed method can effectively extract the input correlation.

\subsection{Model Description}
\label{sec4a}

In this model, we sample protons one by one, each ($\varphi_i$) from a specific azimuthal probability distribution 
determined by the positions of all previously generated protons. The first proton follows a uniform distribution in the interval $[0, 2\pi)$. The PDF for the second proton is a combination of a uniform distribution and a Gaussian centered at $\varphi_1$, $P(\varphi_2) \propto 1 + A e^{-\frac{1}{2}\left(\frac{\varphi_2-\varphi_1}{\sigma}\right)^2}$, with appropriate normalization. The third proton has $P(\varphi_3) \propto \left[1 + A e^{-\frac{1}{2}\left(\frac{\varphi_3-\varphi_1}{\sigma}\right)^2}\right]\left[1 + A e^{-\frac{1}{2}\left(\frac{\varphi_3-\varphi_2}{\sigma}\right)^2}\right]$. This process continues, and the PDF for the $n$-th proton receives a contribution from all previous protons:
\begin{equation}
P(\varphi_{n>2}) \propto \prod_{i=1}^{n-1} {\left[1 + A e^{-\frac{1}{2}\left(\frac{\varphi_n-\varphi_i}{\sigma}\right)^2}\right]}.
\label{eq:gaus_model}
\end{equation}
Since Gaussian distributions extend from $-\infty$ to $\infty$, $\varphi_i$ could fall outside the $[0, 2\pi)$ range. In that case, we fold it back into $[0, 2\pi)$.

%The $\sigma$ parameter is strictly greater than zero. Small $\sigma$ values produce a sharp and thin peak, resulting in large correlations in the vicinity of the track but having little influence at further azimuthal distances. A large $\sigma$ produces a weak but far reaching correlation. 
The two parameters in this Gaussian correlation model, $A$ and $\sigma$, represent the strength and the range, respectively, of the correlation among protons.
Parameter $A$ can take either positive or negative values,
corresponding to positive (clustering) or negative   (repulsion) correlations. The range of $A$, $[-1, +\infty)$, is asymmetric about zero. $A$ being $-1$ prevents any proton from being placed exactly on top of a pre-existing proton. Conversely, $A$ must approach $+\infty$ to ensure each proton is placed directly atop the first one. This asymmetry implies that simulation results with positive and negative values of $A$ should not be compared on the same footing, except for small magnitudes of  $A$.

\subsection{\texorpdfstring{$\Delta \sigma^2$ }{}Analysis}
\label{sec:gaus_cor_model_analysis}

Utilizing the Gaussian correlation model, we create proton events with $N$ ranging from 2 to 80, encompassing the maximum multiplicity observed in the AMPT data. Subsequently, we analyze these events using the same approach as conducted in Sec.~\ref{sec:model_analysis}. We systematically explore different combinations of $A$ and $\sigma$ parameters to understand the observable's behavior across the model's parameter space. In this simulation, there is no need for correction regarding either mixed events or elliptic flow.
We use 72 partition samples per event.

\begin{figure}[t]
\includegraphics[width=\linewidth]{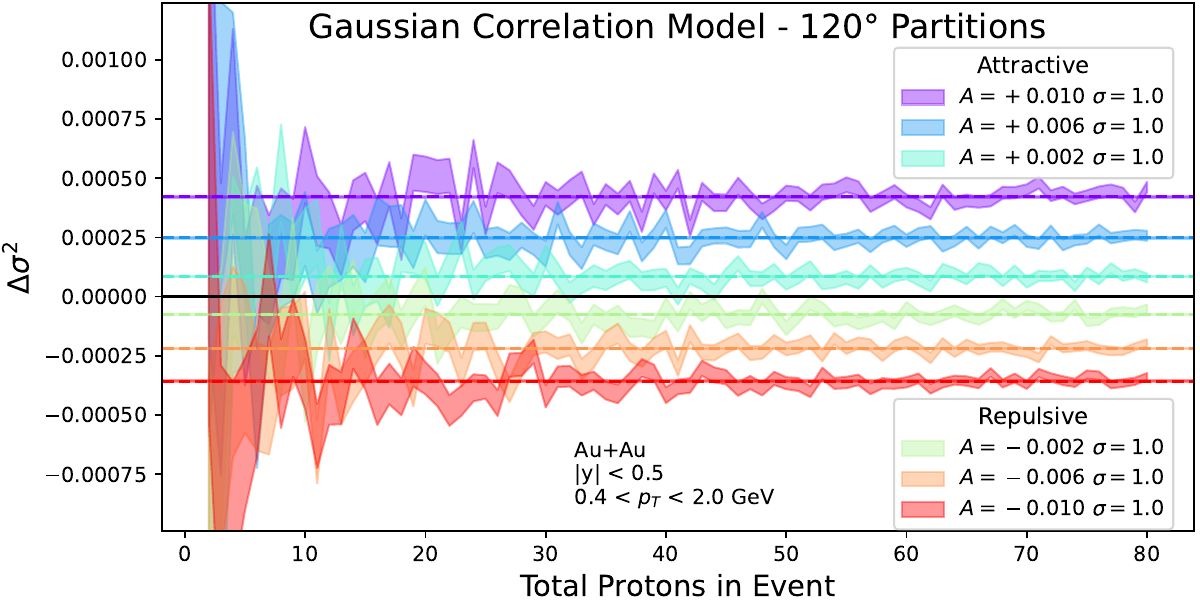}
\caption{Simulations of $\Delta\sigma^2(N_{120})$ as a function of $N$ from the Gaussian correlation model. The input parameters are $A = \pm0.002$, $\pm0.006$, and $\pm0.01$, and $\sigma=1$.}
\label{fig:dsig_vs_tprotons_sim_example}
\end{figure}

Figure~\ref{fig:dsig_vs_tprotons_sim_example} shows the simulations of $\Delta\sigma^2(N_{120})$ as a function of $N$ from the Gaussian correlation model, with the input parameters $A = \pm0.002$, $\pm0.006$, and $\pm0.01$, and $\sigma=1$.  All attractive sets with positive $A$ produce positive $\Delta\sigma^2$ values, whereas all repulsive sets with negative $A$ render negative $\Delta\sigma^2$ values. In addition, larger $|A|$ values lead to larger magnitudes of $\Delta\sigma^2$. Thus, $\Delta\sigma^2(N_{120})$ appears to be correlated with the strength of the proton correlation. Similar to Fig.~\ref{fig:dsig2_vs_tprotons_energies}, $\Delta\sigma^2$ remains constant against $N$. We will use $\langle \Delta\sigma^2 \rangle$, averaged over $N$, in the following analyses.

\begin{figure}[b]
\includegraphics[width=\linewidth]{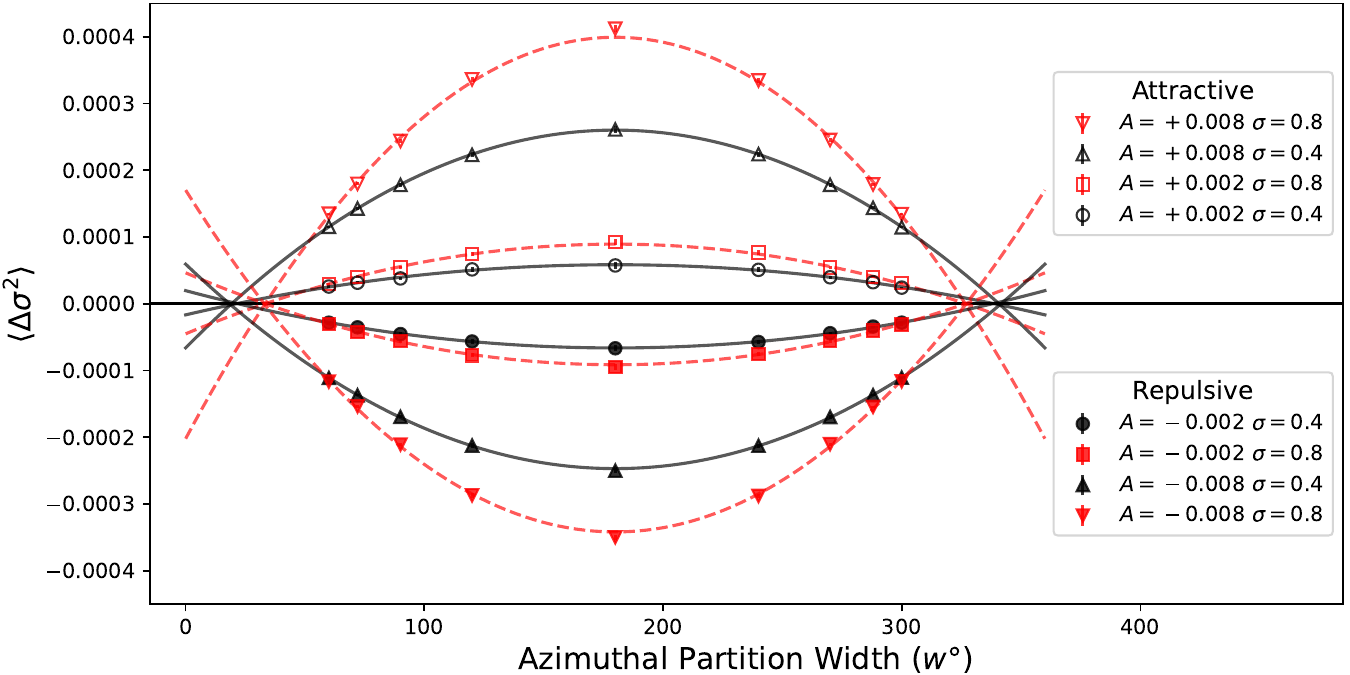}
\caption{$\langle \Delta\sigma^2 \rangle$, obtained by averaging over $N$, vs the width of the azimuthal partition, $w$, from the Gaussian correlation model. The input parameters are $A = \pm0.002$ and  $\pm0.008$, and $\sigma=0.4$ and 0.8.}
\label{fig:dsig_vs_div_sim_sigma1_example}
\end{figure}

%To better investigate the relation between $\langle \Delta\sigma^2 \rangle$ of Figure~\ref{fig:dsig_vs_tprotons_sim_example} and the $A$ parameter, we plot $\langle \Delta\sigma^2 \rangle$ vs $A$ in Figure~\ref{fig:dsigma_vs_a_sim_example}. From the superposed fit lines, we can see that there is a linear relationship between $\langle \Delta\sigma^2 \rangle$ and the $A$ parameter for a given $\sigma$ value in both the attractive and repulsive simulations. 
%This linear relation is dependent upon the value of $\sigma$ used in the simulation. % Not a critical observation. In addition, extrapolating to $A=0$ gives a $\langle \Delta\sigma^2 \rangle$ intercept of zero as expected since a simulation with $A=0$ distributes tracks randomly about the azimuth.

%\begin{figure}
%\includegraphics[width=\linewidth]{figures/gaus_cor_analysis/Dsigma_Avgs_vs_Simulation_Amplitude.pdf}
%\caption{$\langle \Delta\sigma^2 \rangle$ plotted on the y-axis against the amplitude ($A$) simulation parameter on the x-axis. Both attractive and repulsive simulations are run with $\sigma \in [0.8, 1.0, 1.2]$ and $A \in \pm[0.002, 0.004, 0.006, 0.008, 0.01]$. A linear fit is superposed for each of the six sets shown, demonstrating the linearity between $\langle \Delta\sigma^2 \rangle$ and $A$.}
%\label{fig:dsigma_vs_a_sim_example}
%\end{figure}

Figure~\ref{fig:dsig_vs_div_sim_sigma1_example} shows the simulations of $\langle \Delta\sigma^2 \rangle$ as a function of $w$ from the Gaussian correlation model with input parameters $A = \pm0.002$ and  $\pm0.008$, and $\sigma=0.4$ and 0.8. With the same $w$, the magnitude of $\langle \Delta\sigma^2 \rangle$ increases with $|A|$, regardless of whether the proton interaction is attractive or repulsive. These trends can be reasonably well described by the quadratic function in Eq.~(\ref{eq:quad_fit_ampt}), where the sign of the obtained $b$ parameter matches the sign of the input $A$, and a larger $|A|$ corresponds to a larger $|b|$. 
The fit curves with the same $\sigma$ input seem to intersect the $x$-axis around the same location in Fig.~\ref{fig:dsig_vs_div_sim_sigma1_example}, and therefore, they have similar $z$ values.

%The fit is parameterized such that the curvature, $z$, is expressed in units of $w$. The $b$ parameter corresponds to the maximum magnitude of the fit, 
%what we will call the Amplitude parameter.
%what we will call the baseline. The $z$ parameter, which we refer to as the zeros of the fit, corresponds to the distance in degrees between $w=180^\circ$ and the point at which the fit intersects the x-axis. It is inversely related to the curvature of the quadratic.  This is found to be true for simulations with other $\sigma$ values, but the intersection point changes with $\sigma$. In addition, the intersection can be different between the attractive and repulsive simulations. Focusing on repulsive correlations, Figure~\ref{fig:dsig_vs_div_sim_sigma1_example} shows that simulations with the same value of $\sigma$ intersect the x-axis at the same place and that this intersection changes with $\sigma$.

%\begin{figure}
%\includegraphics[width=\linewidth]{figures/gaus_cor_analysis/dsig_vs_width_sigma04_12_example.pdf}
%\caption{$\langle \Delta\sigma^2 \rangle$ plotted against the width of the azimuthal partitions used. Simulation Amplitudes of $A \in [-0.006, -0.01]$ are shown with $\sigma \in [0.4, 1.2]$.}
%\label{fig:dsig_vs_div_sim_sigma04_12_example}
%\end{figure}

\begin{figure}
\includegraphics[width=\linewidth]{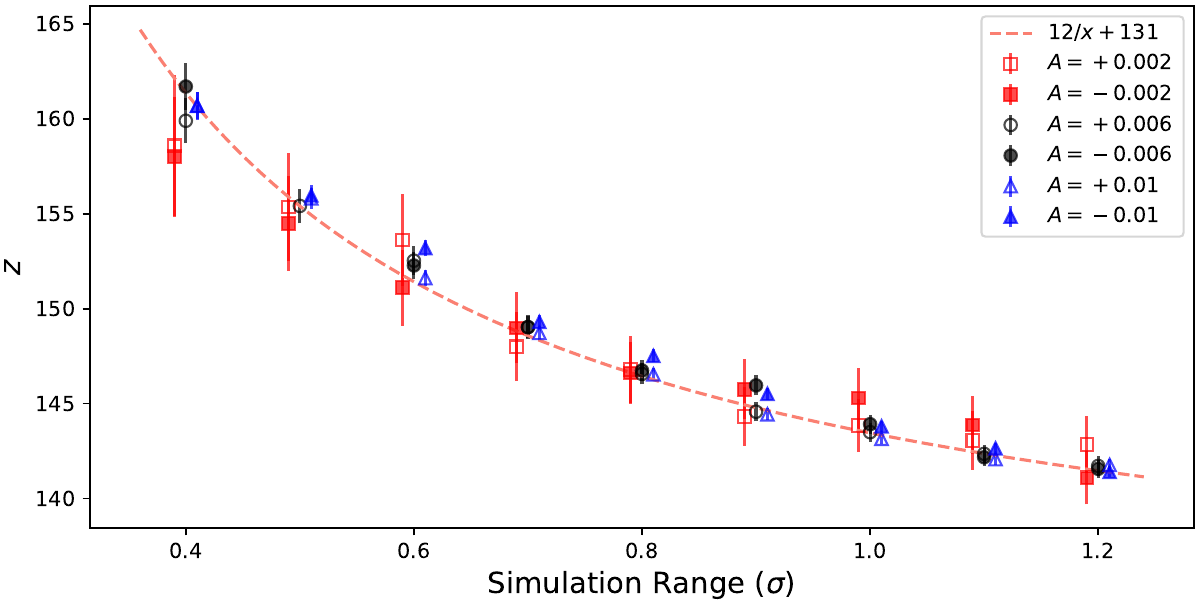}
\caption{The inverse curvature $z$ extracted from the $w$ dependence of $\langle \Delta \sigma^2 \rangle$ using the quadratic fit in Eq.~(\ref{eq:quad_fit_ampt})  as a function of input parameter $\sigma$ from the Gaussian correlation model.
The input parameter $A$ is chosen to be $\pm0.002$, $\pm0.006$, and $\pm0.01$.
The dashed curve represents a simple function, $z=12/\sigma+131$,  describing the relation between $z$ and $\sigma$. Some markers have been slightly shifted along the $x$-axis to improve clarity.
}
\label{fig:quad_fit_z_vs_spread}
\end{figure}

Figure~\ref{fig:quad_fit_z_vs_spread} shows the inverse curvature $z$ extracted from the $w$ dependence of $\langle \Delta \sigma^2 \rangle$ using the quadratic fit in Eq.~(\ref{eq:quad_fit_ampt})  as a function of input parameter $\sigma$ from the Gaussian correlation model, with
input parameter $A = \pm0.002$, $\pm0.006$, and $\pm0.01$.
Note that the units of $z$ and $\sigma$ are degree and radian, respectively. The extracted $z$ seems to be solely dependent on the $\sigma$ value, independent of input parameter $A$. As illustrated by the dashed curve, $z$ follows a simple function of $\sigma$, $z = 12/\sigma + 131$.
This supports the idea that $z$ holds a meaningful physical interpretation, reflecting the correlation range.
According to the realistic AMPT simulations in Fig.~\ref{fig:z_vs_refmult}, the $z$ values typically fall at or below $145^\circ$, corresponding to $\sigma\ge0.8$ in Fig.~\ref{fig:quad_fit_z_vs_spread}.

\begin{figure}
\includegraphics[width=\linewidth]{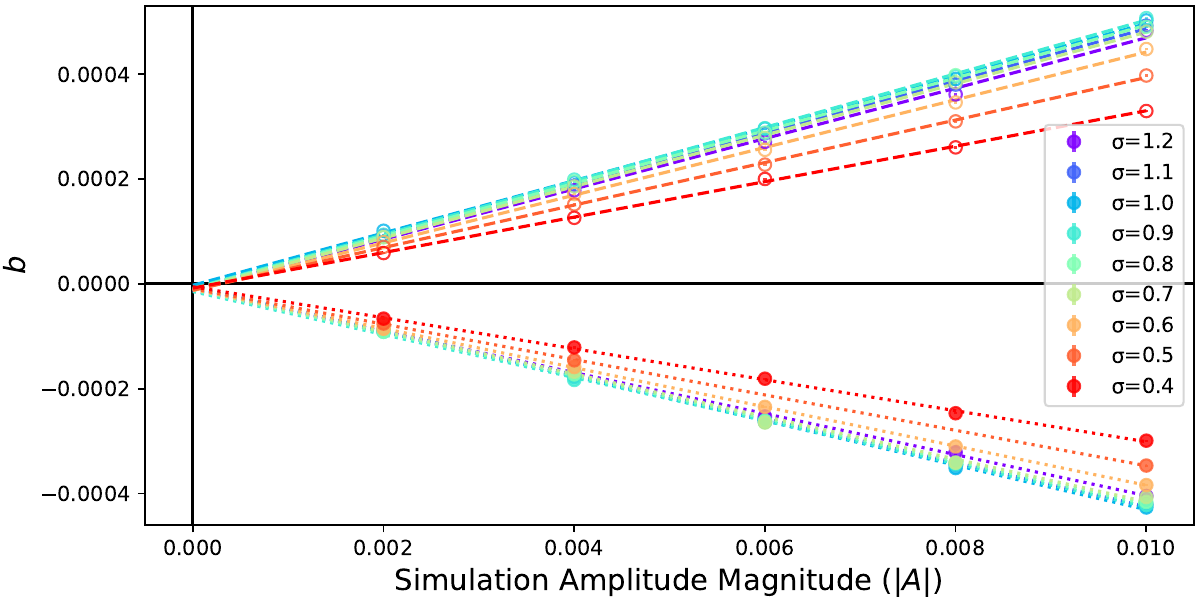}
\caption{The general scale $b$ extracted from the $w$ dependence of $\langle \Delta \sigma^2 \rangle$ using the quadratic fit in Eq.~(\ref{eq:quad_fit_ampt})  as a function of input parameter $|A|$ from the Gaussian correlation model.
The positive (negative) $b$ values correspond to attractive (repulsive) interactions with positive (negative) $A$.
The input parameter $\sigma$ varies from 0.4 to 1.2. The dashed lines represent linear functions passing through the point of origin.}
\label{fig:quad_fit_b_vs_A}
\end{figure}

Figure~\ref{fig:quad_fit_b_vs_A} presents the general scale $b$  extracted from the $w$ dependence of $\langle \Delta \sigma^2 \rangle$ using the quadratic fit in Eq.~(\ref{eq:quad_fit_ampt})  as a function of input parameter $|A|$ from the Gaussian correlation model, with $\sigma$ ranging from 0.4 to 1.2. The sign of $b$ corresponds to positive or negative $A$, indicating attractive or repulsive proton interactions. In cases where $\sigma \geq 0.8$, $b$ follows a linear function of $A$ passing through $(0, 0)$, regardless of the magnitude of $\sigma$.
Hence, in realistic scenarios, $b$ serves as an indicator of the correlation strength in this model.
When $\sigma\le0.8$, the $b$ values deviate from the universal trend, but the linear relation between $b$ and $A$ still holds. If real data yield $z$ values corresponding to small $\sigma$, the measured $b$ can still be employed to determine the correlation strength, and its sign indicates whether the interaction is attractive or repulsive.

One concern regarding the Gaussian correlation model is the 
sequential placement of protons.
While this procedure mimics the interaction between protons, the protons may not be treated equally. As each proton is generated according to the positions of all previous ones, protons placed earlier appear to be influenced by fewer other protons. 
However, this effect is expected to be negligible for two reasons.
Firstly, the two angles in the Gaussian distribution in Eq.~(\ref{eq:gaus_model}) are mathematically symmetric, indicating that the implemented interaction between two protons is mutual.
Secondly, the $\Delta\sigma^2$ observable randomly samples and resamples all protons without any specific sequence or bias.

%\red{Another consideration pertains to integrating this model with other types of correlations. For instance, if there is a void in the detector acceptance, causing zero probability of producing protons in a specific azimuthal region, the Gaussian correlation model will treat the void as an area to avoid (fill) if the input interaction is attractive (repulsive). This effectively introduces a coupling between the efficiency PDF and this model. A potential solution is to generate protons with the model first and then reject some of them based on efficiency. However, this handling may face challenges in ensuring the proper number of protons produced in a given event. While such couplings are typically minor, with some interpretation, this model can still be valuable when combined with other sources of correlations.}
%\blue{We don't actually run the model with anything else in this paper. The warning is included in my thesis where I do test with other models. Maybe we can remove the paragraph? No strong opinion. DN}

\section{Conclusion}
\label{sec6}
To explore the nature of the phase transition in heavy-ion collisions, we advocate a new approach to investigate proton multiplicity fluctuations in azimuthal partitions. We introduce a $\Delta \sigma^2$ observable to quantify the azimuthal clustering of final-state protons. The robustness of the measurement against detector inefficiency is ensured through the normalization scheme and the mixed-event correction. Additionally, the correction for elliptic flow eliminates the contribution of collective motion to the clustering signal. Leveraging the event resampling technique allows for optimal utilization of available statistics.

We have applied the proposed methodology to simulations using the AMPT and the MUSIC+FIST models for Au+Au collisions at center-of-mass energy ranging from 7.7 to 62.4 GeV. The resulting $\Delta \sigma^2$ values remain independent of the total proton multiplicity ($N$), validating the effectiveness of the normalization scheme.
The discrepancy in $\Delta \sigma^2$ between the two versions of MUSIC+FIST highlights the observable's sensitivity to the excluded-volume effect.
The variation in AMPT results with collision energy and event multiplicity demonstrate general features due to global momentum conservation.

%A weak repulsion was measured in the MUSIC+FIST model while a stronger, energy-dependent repulsion was measured in AMPT. It was found that this repulsive effect is best described as a function of the event multiplicity, in which the magnitude increases quickly as event multiplicity decreases. We hypothesize that this strong event multiplicity dependence is a consequence of momentum conservation.

With both models, we have also investigated the dependence of 
$\langle \Delta\sigma^2 \rangle$, averaged over $N$, on the azimuthal partition width $w$. We characterize the $w$ dependence with a quadratic function, utilizing parameters $b$ and $z$ to quantify the general scale and the inverse curvature, respectively.
With a Gaussian correlation model, we illustrate that the extracted $b$ and $z$ values reflect the strength and the range, respectively, of the input interaction between protons.

%While the Gaussian correlation model is rather simple, our tests of the $\langle \Delta\sigma^2 \rangle$ observable motivates performing similar quadratic fits to A+A collision data and simulated data from dynamical models and interpreting the $b$ parameter of these fits as a proxy for correlation strength and the $z$ parameter as a proxy for the correlation range.

The application of the proposed approach to the RHIC BES data will provide new insight into the nature of the quark-hadron phase transition in the QCD phase diagram. In particular, measurements of the innovative observable $ \Delta\sigma^2$ will allow us to probe the clustering phenomenon among protons. 
While addressing the overall scale of $ \Delta\sigma^2$ necessitates further evaluation of the momentum conservation effect, the inverse curvature shows little dependence on the scale magnitude and reflects mostly the correlation width in the Gaussian correlation model. The correlation width in azimuth could be linked to the underlying physics of the phase transition, making its beam-energy dependence of great experimental interest.

\begin{acknowledgments}
{The authors thank Zhiwan Xu, Yunshan Cheng, Thomas Marshall, Aditya Dash, and Xiatong Wu for many fruitful discussions. Z. J., G. W., and H. H. are supported
by the U.S. Department of Energy under Grant No. DE-FG02-88ER40424 and by the National Natural Science Foundation of China under Contract No.1835002. We acknoledge the support from the Office of Nuclear Physics in the Office of Science of the Department of Energy and the Commissariat `a l’Energie Atomique (France).
}
\end{acknowledgments}

\appendix

\addcontentsline{toc}{section}{\appendixname{}} % Add the appendix title to the table of contents

\section{Impact of Momentum Conservation}
\label{sec:mom_cons_model}

The conservation of momentum in each event is expected to induce a negative correlation between particles. In an extreme case where only two particles are produced back to back, their momenta are perfectly anti-correlated. If more particles emerge in the final state, each becomes less crucial for conserving the total momentum. Consequently, as the number of particles approaches infinity, the anti-correlation between them vanishes. While we comprehend the behavior in extreme cases, the precise functional form of this effect remains elusive.

We have employed a simple model to examine the impact of momentum conservation on the $\Delta \sigma^2$ observable. In each event, we generate $M$ particles, with each momentum component of each particle randomly sampled from a uniform distribution within $(0, 2)$ GeV/$c$. We calculate the total momentum of the event as the sum of individual momenta. Each particle is then slightly rotated away from this total momentum, the rotation amount being proportional to the particle's momentum magnitude and the event's total momentum magnitude. After all particles undergo these slight rotations, we recalculate the total momentum, and the process is repeated. As the rotations diminish with each iteration, the total momentum converges to zero. 
We then randomly label
40\% of these particles as protons, and further apply the kinematic selection with $|y|<0.5$ and  $p<2$ GeV/$c$. 

\begin{figure}[bthp]
\centering
\includegraphics[width=\linewidth]{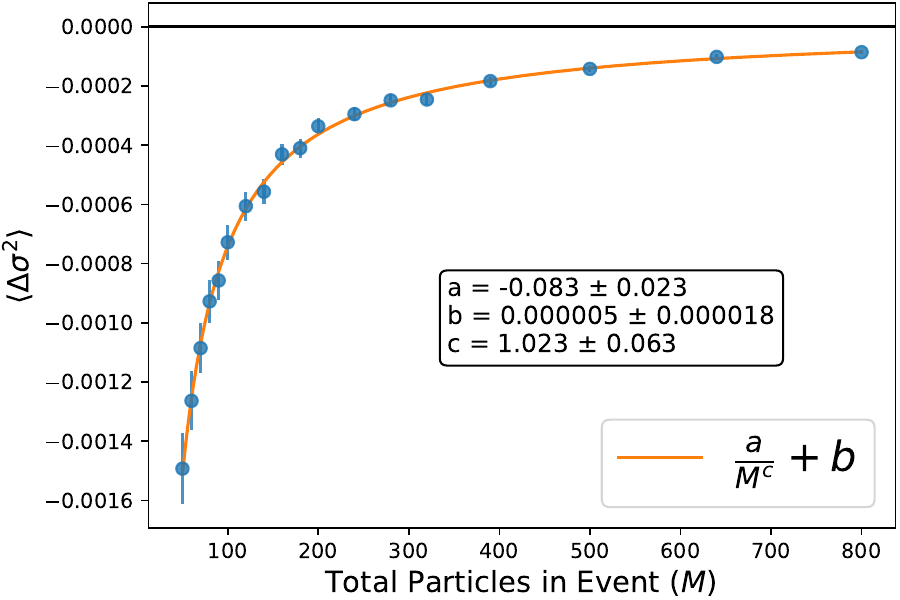}
\caption{Proton $\langle \Delta \sigma^2 \rangle$, averaged over $N$, as a function of  total particle number, $M$, from a simple model that conserves total momentum. The fit function suggests that a simple form of $\propto 1/M$ can well describe the trend.}
\label{fig:p_cons_model_fit}
\end{figure}

Figure~\ref{fig:p_cons_model_fit} displays proton $\langle \Delta \sigma^2 \rangle$, averaged over $N$, as a function of $M$ from the model simulation. As anticipated, the effect of momentum conservation is evident in notably negative $\langle \Delta \sigma^2 \rangle$ values, which become more negative for events with lower multiplicities. We fit this trend with a power law function plus a constant, and the best fit suggests a $1/M$ dependence without an offset. This trend mirrors observations in the AMPT simulations. The straightforward functional relationship indicates that the impact of momentum conservation on the $\langle \Delta \sigma^2 \rangle$ observable may be analytically manageable, implying the potential for a correction procedure.

While the simulation results are enticing, we note that the momentum conservation model presented here is oversimplified. Several factors, such as particle spectra, the proton fraction, and the particle momentum distribution may impact the $\langle \Delta \sigma^2 \rangle$ values. Additionally, the model lacks many physics aspects involved in heavy-ion collisions, and it is beyond the scope of this study to construct a realistic model for such collisions. 

\bibliographystyle{apsrev4-2-dylan}
\bibliography{references}

\end{document}